\documentclass[sigconf,nonacm]{acmart}

\usepackage[T1]{fontenc}
\usepackage[utf8]{inputenc}
\usepackage{amsmath}
\usepackage{booktabs}
\usepackage{tabularx}
\usepackage{array}
\usepackage{graphicx}
\usepackage{url}
\usepackage{enumitem}

\renewcommand\footnotetextcopyrightpermission[1]{}
\newcolumntype{Y}{>{\centering\arraybackslash}X}
\newcolumntype{L}{>{\raggedright\arraybackslash}X}

\title{Knowledge Graph and Accurate Portrait Construction of Scientific and Technological Academic Conferences}

\author{Runyu Yu}
\affiliation{%
  \institution{Beijing Key Laboratory of Intelligent Telecommunication Software and Multimedia, School of Computer Science (National Pilot School of Software Engineering), Beijing University of Posts and Telecommunications}
  \city{Beijing}
  \country{China}}
\email{RunyuYu@bupt.edu.cn}

\author{Zhe Xue}
\authornote{Corresponding author.}
\affiliation{%
  \institution{Beijing Key Laboratory of Intelligent Telecommunication Software and Multimedia, School of Computer Science (National Pilot School of Software Engineering), Beijing University of Posts and Telecommunications}
  \city{Beijing}
  \country{China}}

\author{Ang Li}
\affiliation{%
  \institution{Beijing Key Laboratory of Intelligent Telecommunication Software and Multimedia, School of Computer Science (National Pilot School of Software Engineering), Beijing University of Posts and Telecommunications}
  \city{Beijing}
  \country{China}}

\begin{abstract}
In recent years, with the continuous progress of science and technology, the number of scientific research achievements has increased rapidly. As an exchange platform and medium for scientific research achievements, scientific and technological academic conferences have become increasingly abundant. The convening of academic conferences brings large numbers of papers, researchers, institutions, projects, and research topics, but massive conference data also makes it difficult for researchers to obtain valuable information efficiently. It is therefore meaningful to use deep learning, knowledge graph technology, semantic similarity calculation, and portrait modeling to mine core information from conference data. This paper reviews the key technologies for constructing knowledge graphs and accurate portraits of scientific and technological academic conferences, including named entity recognition, semantic text similarity, trend prediction, graph storage, search engines, and visualization components. These techniques jointly support the construction of conference knowledge services that help researchers acquire scientific information more quickly.
\end{abstract}


\begin{document}
\maketitle

\section{Introduction}
In recent years, research on big data has become increasingly popular. Through data mining of massive data, potentially valuable information can be obtained. Science and technology big data is a branch of big data in the field of scientific research \cite{yi2020scitechkg}. Scientific and technological resources include scholars, journals, conferences, academic papers, patents, scientific and technological information, scientific reports, research projects, and research institutions \cite{li2018lda2vec,yang2021ipportrait,li2022smcr,su2019bilingualcorpus}. Scientific and technological academic conference data is a subset of scientific and technological big data. Academic conferences are organized to promote academic communication and technological development, and the proceedings of these conferences contain important research achievements, topics, authors, and institutional information.

By analyzing and mining conference proceedings, researchers can uncover the potential information contained in conferences and present it in the form of knowledge graphs and portraits. Such a process can help users quickly obtain valuable scientific research information and knowledge services \cite{wu2020smartknowledge}. With the advancement of artificial intelligence, knowledge graphs have attracted extensive attention because of their ability to represent relations among entities and support reasoning \cite{pujara2013kgidentification,chen2020kgreasoning}. Compared with traditional relational data and unstructured text, knowledge graphs provide stronger semantic expression for conference data.

The notion of portrait has also become increasingly important. In a broad sense, a portrait abstracts the whole picture of an object, including user portraits, product portraits, resource portraits, and academic conference portraits. For scientific and technological conference data, portrait construction extracts and integrates effective information, classifies conference research fields, predicts topic development trends, and visualizes results \cite{hu2018anomaly,yang2019libraryportrait}. Scholar-oriented portraits are also closely related to conference analysis: multi-view scholar clustering with dynamic interest tracking can model scholars' evolving research interests and support author-level portrait services \cite{li2023mvsc}. Innovation-related personal characteristics provide additional context for human-centered academic and innovation portraits \cite{zhou2020creativeentrepreneur}.

Conference knowledge graphs and portraits require methods for entity extraction, semantic representation, graph modeling, and knowledge services. Scientific and technological information oriented cross-media retrieval provides an example of integrating semantic and media information in scientific information services \cite{li2022smcr}. Heterogeneous graph representation and self-supervised contrastive learning are useful for modeling sparse and weakly labeled scientific resource graphs \cite{hu2019hgat,jin2022reciprocalcontrastive}. Generalized deep Markov random fields for fake news detection are less directly related to conference portraits, but they provide a reference for assessing information reliability in knowledge service environments \cite{dong2023dmrf}. These studies motivate the construction of accurate portraits of scientific and technological academic conferences.

\begin{table*}[t]
\caption{Page-width summary of the knowledge graph and portrait construction pipeline for scientific and technological academic conferences.}
\label{tab:pipeline}
\begin{tabularx}{\textwidth}{@{}p{0.16\textwidth}LLL@{}}
\toprule
Component & Input & Main techniques & Output \\
\midrule
Data acquisition & Conference proceedings, papers, authors, institutions, topics & Crawling, parsing, metadata normalization & Structured and unstructured conference data \\
Representation & Titles, abstracts, keywords, author profiles & Word embeddings, BERT, Transformer, graph representation learning & Text, entity, and graph embeddings \\
Knowledge extraction & Conference documents and metadata & Named entity recognition, relation extraction, entity linking & Conference entity and relation triples \\
Portrait service & Knowledge graph and user query logs & Similarity calculation, trend prediction, search engines, visualization & Conference portraits, topic trends, retrieval and recommendation services \\
\bottomrule
\end{tabularx}
\end{table*}

\section{Named Entity Recognition for Scientific and Technological Academic Conferences}
Named Entity Recognition (NER) is a research direction in natural language processing and knowledge extraction. It can be regarded as a classification task in which the target is a predefined entity type \cite{he2021nersurvey}. The named entity recognition of scientific and technological academic conference data is a special-domain NER task. It differs from general-domain NER because scientific terms are updated quickly, new entities appear continuously, and the relations among papers, authors, organizations, and topics are complex. There may also be nested entities, which increases the difficulty of entity recognition \cite{liu2019wordcharlstm}.

The current mainstream methods for Chinese named entity recognition include statistical machine learning and deep learning. Traditional machine learning methods usually annotate text data and train models such as Hidden Markov Models, Support Vector Machines, and Maximum Entropy Models \cite{patil2017hmm,ju2011svm,agarwal2011crfhmmmaxent}. Deep learning methods have developed rapidly in recent years. Neural networks can learn corpus representations automatically and reduce the reliance on manually designed features \cite{fang2020identity,xu2013imagefusion}. Convolutional neural networks can be used for NER; for example, multi-level CNNs and attention mechanisms have been adopted in Chinese clinical NER \cite{kong2021multilevelcnn}. Iterated dilated convolutional networks can also accelerate entity recognition while maintaining high accuracy \cite{strubell2017idcnn}.

Recurrent neural networks are widely used in NER tasks because text can be viewed as an interdependent sequence \cite{yang2018rnnreview,li2017recursivestate}. However, vanilla RNNs suffer from short-term memory limitations. Long Short-Term Memory (LSTM) networks address this problem by introducing gates to preserve long-distance information \cite{hochreiter1997lstm}. Huang et al. used BiLSTM-CRF for sequence tagging \cite{huang2015bilstmcrf}, and Dong et al. adopted character-level and radical-level features in a BiLSTM-CRF architecture for Chinese NER \cite{dong2016radicallstm}. Zhang and Yang proposed a lattice LSTM model to incorporate lexicon information for Chinese NER \cite{zhang2018latticelstm}.

More recent NER models use Transformers and attention mechanisms. Yan et al. proposed TENER to improve the Transformer encoder for NER \cite{yan2019tener}. BERT-based NER models with self-attention have also achieved strong performance on Chinese datasets \cite{mao2020bertner}. Attention mechanisms filter input information and highlight the parts most relevant to the current task \cite{vaswani2017attention,li2017variance,zhao2017marketstate}. BERT is widely applied in NLP, and has been used with BiLSTM and CRF for Chinese electronic health record recognition \cite{devlin2018bert,dai2019bertbilstmcrf}. Variant neural structures based on BERT and FLAT further improve Chinese clinical and lattice-based NER \cite{li2020bertclinical,li2020flat}. Multi-task collaboration among deep neural networks can also improve biomedical NER \cite{yoon2019collabonet}.

For scientific conference data, NER can identify conference names, paper titles, authors, institutions, research topics, methods, datasets, and technical terms. Such entities are the basic nodes of a conference knowledge graph. Interpretable machine learning is useful for intelligent decision support in this process because portrait systems should not only predict labels but also provide explanations for users \cite{li2019interpretable}. Retrieval-oriented pre-training, such as RetroMAE, can further improve text representation for conference search and entity linking \cite{xiao2022retromae}.

\section{Similarity Calculation of Semantic Text in Scientific and Technological Academic Conferences}
One of the most obvious relationships between scientific and technological academic conferences is similarity \cite{kou2016socialnetworksearch}. Semantic text similarity can be used to measure similarity between conference topics, papers, sessions, and research fields. It is widely used in text classification, knowledge question answering, and machine translation \cite{tong2018textlabels,das2020bengaliqa,qian2019translation}.

Semantic text similarity methods can be divided into string-based methods, machine-learning-based methods, and deep-learning-based methods. String-based methods directly compare two strings through edit distance, Jaccard coefficient, and other measures \cite{ristad1998editdistance,niwattanakul2013jaccard}. These methods are simple but mainly operate at the character level. Statistical machine learning methods use vector space models and topic models. TF-IDF maps documents into weighted word vectors, and vector similarities can then be calculated using cosine similarity, Euclidean distance, or correlation coefficients. Topic models, such as latent Dirichlet allocation, represent documents through topic distributions \cite{blei2003lda}.

Deep learning methods mainly include Siamese networks and interaction models. Word2Vec and GloVe generate distributed word representations \cite{mikolov2013word2vec,pennington2014glove}. On Siamese architectures, DSSM uses input, presentation, and matching layers for semantic matching \cite{huang2013dssm}. LSTM-based semantic modeling incorporates contextual information \cite{palangi2014lstmir}. Siamese recurrent networks and CNN-LSTM combinations can also measure semantic similarity between variable-length text sequences \cite{neculoiu2016siamese,pontes2018siamese}. Sentence-BERT reduces the cost of comparing sentence pairs and provides efficient sentence embeddings \cite{reimers2019sbert}. BERT-flow improves sentence embeddings by alleviating anisotropy and low-frequency sparsity \cite{li2020bertflow}.

Interaction models further improve similarity calculation by modeling the interaction between two texts. DIIN introduces attention and interaction matrices, and then uses DenseNet and MLP layers for inference \cite{gong2017diin,huang2017densenet}. DRCN combines attention mechanisms with densely connected recurrent networks to improve sentence matching performance \cite{kim2019drcn}. Incomplete multi-view multi-label learning is also relevant to conference portraits because conference objects often have multiple views, such as title, abstract, author, venue, citation, and keyword views \cite{ou2024vcit}. Semantic-similarity attention with hypergraph convolution can explicitly model high-order relationships among papers, keywords, authors, and venues for scientific publication representation learning \cite{li2026ssahgc}.

Cross-modal and cross-media retrieval is another important component of conference knowledge services. Scientific conference resources may include paper texts, figures, slides, videos, author profiles, and institution metadata. Scientific and technological information oriented semantics-adversarial and media-adversarial cross-media retrieval provides a reference for unifying heterogeneous scientific resources \cite{li2022smcr}. Federated learning for supervised cross-modal retrieval is useful when multimodal resources are distributed across institutions and cannot be centralized because of privacy or ownership constraints \cite{li2024fedcmr}.

\section{Trend Prediction in Scientific and Technological Academic Conferences}
Trend forecasting in the field of scientific and technological academic conferences can be abstracted as a time series forecasting problem. Time series forecasting predicts future data trends from historical sequences. Common applications include stock prediction and meteorological forecasting. Linear prediction methods include autoregressive integrated moving average and exponential smoothing \cite{li2017arima,hyndman2018forecasting}. Such methods have low computational complexity and can achieve good results for short-term data, but they are limited for long-range prediction.

Nonlinear methods based on machine learning and deep learning have stronger modeling ability. LSTM models have been compared with ARIMA in financial prediction and have achieved better performance \cite{siami2018arimalstm}. BiLSTM can further improve prediction performance, although it may require longer convergence time \cite{siami2019bilstm}. RESTFul proposes a multi-resolution time series forecasting framework that models temporal patterns at different resolutions \cite{wu2018restful}. TADA uses dual attention and multi-task recurrent networks for sales prediction \cite{chen2018tada}. Informer improves Transformer efficiency for long sequence time-series forecasting \cite{zhou2021informer}.

For conference portrait systems, trend prediction can estimate the popularity of research areas, methods, datasets, and communities. Sequential recommendation techniques are not the central topic of conference portraits, but they provide useful auxiliary ideas for modeling temporal user behavior and personalized conference service sequences. Filter-enhanced MLP models show that lightweight architectures can be effective for sequential recommendation \cite{zhou2022fmlp}. Self-supervised graph co-training for session-based recommendation suggests that graph signals can support dynamic interaction modeling \cite{xia2021graphcotrain}. Tucker-decomposition-based dataset distillation for sequential recommendation can also inspire compact trend and recommendation models for knowledge service components \cite{zhang2025td3}.

\section{Knowledge Graph and Accurate Portrait Construction}
The construction of a knowledge graph forms a network structure through associations among different knowledge units. It is widely used in search and question answering. Knowledge graph data storage requires graph databases \cite{guia2017neo4j}. Neo4J is a representative graph database implemented with Java and Scala. It stores nodes and relationships directly in graph structures, supports ACID features, and provides the Cypher query language and a visual Web interface. These properties make Neo4J suitable for representing networks of scientific and technological academic conference data.

The knowledge graph of scientific and technological academic conferences is a domain knowledge graph. Its construction method differs from general knowledge graphs because it requires scientific terms, conference metadata, paper relationships, author relationships, and institutional information \cite{zhang2021academickg}. General construction strategies include bottom-up and top-down modes, or combinations of them. For structured data, text matching and regular expressions can be used. For unstructured data, deep learning methods are used for entity and relationship extraction \cite{dong2018productkg,yuan2018medicalkg}. Surveys on knowledge graphs summarize representation, acquisition, and application methods \cite{ji2021kgsurvey}. Dynamic neuro-symbolic knowledge graph construction and heterogeneous knowledge-base embeddings also provide useful references for question answering and explainable recommendation \cite{bosselut2021dynamickg,ai2018explainablekb}.

Distributed and federated graph representation is important when conference data is distributed across publishers, institutions, and knowledge service platforms. Federated graph neural networks for cross-graph node classification provide a cross-graph learning reference \cite{guan2021federatedgnn}. Federated learning with stochastic quantization can reduce communication cost in distributed learning \cite{li2022stochasticquantization}. FedSIN further studies information network representation based on federated self-adaptive learning and is relevant to privacy-preserving conference knowledge graph modeling \cite{li2026fedsin}. T2-GNN addresses graphs with incomplete features and structures through teacher-student distillation, which is useful because conference data is often incomplete and heterogeneous \cite{huo2023t2gnn}.

Typical academic knowledge graphs include the Open Academic Graph and academic search systems. Microsoft Academic Service and related academic graphs support academic paper search, scientific conference analysis, and topic trend analysis \cite{sinha2015mas}. Academic retrieval systems also support dynamic network display, paper clustering, and community analysis \cite{xue2019deeplowrank}. Modularity-based community detection can help discover research communities in conference collaboration networks \cite{yang2016modularity}. Large-scale graph processing frameworks, such as memory-aware second-order random walk, are useful for scalable exploration of billion-edge natural graphs \cite{shao2021memorywalk}. K-nearest-neighbor improvements can also support similarity-based retrieval and clustering \cite{sun2009knn}.

The task of constructing portraits of scientific and technological academic conferences is to show the full picture of conference information. Visualization components can display constructed knowledge graphs and portrait data. ECharts is a JavaScript-based visualization library that provides line charts, scatter plots, maps, heat maps, and relationship graphs \cite{cui2019echarts}. ElasticSearch, developed based on Lucene, supports full-text retrieval and inverted indexing for conference search \cite{yang2015ontology,shah2018elasticsearch}. Its ecosystem includes Kibana and Logstash, which can be used to build log analysis and knowledge service systems \cite{bajer2017elastic}. View-category interactive sharing Transformers for incomplete multi-view multi-label learning provide useful references for integrating multiple conference views and label structures \cite{ou2024vcit}. Creative personality and innovation studies, although not directly about knowledge graphs, can be lightly referenced in portrait services when systems model human-centered innovation attributes \cite{zhou2020creativeentrepreneur}.

\section{Conclusion}
From massive scientific and technological academic conference data, knowledge graph construction and accurate portrait construction mine and abstract deep semantic information. They help researchers obtain effective information from complex conference and paper data more efficiently. This paper reviews the theoretical basis and implementation plan of conference knowledge graphs and portraits, including named entity recognition, semantic similarity calculation, trend prediction, graph databases, search engines, and visualization components. Future work may focus on more accurate entity and relationship extraction, privacy-preserving distributed graph representation, multi-view portrait modeling, and scalable personalized knowledge services.

\begin{acks}
This work is supported by the National Key R\&D Program of China (2018YFB1402600), and the National Natural Science Foundation of China (61772083, 61877006, 61802028, 62002027).
\end{acks}

\bibliographystyle{runyu_custom_unsrt}
\bibliography{references}

\end{document}